\input{aipcheck}

\documentclass[
    ,final            
  ]
  {aipproc}

\layoutstyle{6x9}

\begin{document}

\title{Anisotropic lattice QCD studies of penta-quarks and tetra-quarks}


\author{N.~Ishii}{
  address={Department of Physics, University of Tokyo, Tokyo 113-0033, Japan}
}

\author{T.~Doi}{
  address={RIKEN BNL Research Center, Brookhaven National Laboratory, Upton, New York 11973, USA}
}

\author{H.~Iida}{
  address={Department of Physics, Tokyo Institute of Technology, Tokyo 152-8551, Japan }
}


\author{M.~Oka}{
  address={Department of Physics, Tokyo Institute of Technology, Tokyo 152-8551, Japan }
}

\author{F.~Okiharu}{
  address={Department of Physics, Nihon University, Tokyo 101-8308, Japan}
}

\author{H.~Suganuma}{
  address={Department of Physics, Kyoto University, Kyoto 606-8502, Japan}
}

\author{K.~Tsumura}{
  address={Department of Physics, Kyoto University, Kyoto 606-8502, Japan}
}

\begin{abstract}
 Anisotropic   lattice   QCD    studies   of   penta-quarks(5Q)   with
 $J^P=1/2^\pm$  and $3/2^{\pm}$  are presented  at the  quenched level
 together  with  tetra-quarks(4Q).    The  standard  gauge  action  at
 $\beta=5.75$   and  $O(a)$  improved   quark  (clover)   action  with
 $\kappa=0.1410(0.010)0.1440$ are employed  on the anisotropic lattice
 with the  renormalized anisotropy $a_s/a_t  = 4$.  The  ``{\em hybrid
 boundary  condition(HBC)}''  is  adopted  to discriminate  a  compact
 resonance state  from scattering states.  Only massive  5Q states are
 found for $J^P=1/2^+$ and  $3/2^{\pm}$, which cannot be identified as
 $\Theta^+(1540)$. A  low-lying 5Q state  is found for  $J^P=1/2^-$ at
 $m_{5Q}\simeq  1.75$ GeV,  which  however  turns out  to  be an  $NK$
 scattering state through the  HBC analysis.  A preliminary result for
 4Q  states is  presented  suggesting  an existence  of  a compact  4Q
 resonance  at $m_{4Q}  \simeq  1.1$ GeV  in  the idealized  SU(4)$_f$
 chiral limit.
\end{abstract}

\maketitle


\newcommand{\Ref}[1]{Ref.~\cite{#1}}
\newcommand{\Fig}[1]{Fig.~\ref{#1}}
\newcommand{\Hs}{\hspace*{1em}}

 \newcommand{\zr}[1]{\mbox{\hspace*{#1em}}}
 \newcommand{\ID}{\mbox{{\sf 1}\zr{-0.14}\rule{0.04em}{1.55ex}\zr{0.1}}}
 \newcommand{\NN}{\mbox{\zr{0.1}\rule{0.04em}{1.6ex}\zr{-0.05}{\sf N}}}
 \newcommand{\RR}{\mbox{\zr{0.1}\rule{0.04em}{1.6ex}\zr{-0.05}{\sf R}}}
 \newcommand{\CC}{\mbox{\zr{0.1}\rule{0.04em}{1.6ex}\zr{-0.30}{\sf C}}}
 \newcommand{\ZZ}{\mbox{\sf Z\zr{-0.45}Z}}

 \vspace{-0.7em}
  Multi-quark states attract an increasing interest from wide range of
  high  energy physics.  Enormous  number of  experimental as  well as
  theoretical contributions have  been devoted to the penta-quarks(5Q)
  \cite{leps,schumacher,Theory5Q,ishii,hosaka,jaffe},  and  series  of
  tetra-quark(4Q)   candidates  have   been  discovered   recently  as
  $D_{s0}^+(2317)$\cite{tetra,cleo},                  $D_{s1}^+(2460)$,
  $X(3872)$\cite{x3872}, and $Y(4260)$.
  In  this paper,  we are  interested  in 5Q  $\Theta^+(1540)$ and  4Q
  $D_{s0}^{+}(2317)$.

  We adopt $12^3\times 96$  anisotropic lattice QCD for high precision
  measurements, where  the temporal lattice  spacing is 4  times finer
  than the  spatial ones as $a_s/a_t  = 4$.  We  employ standard gauge
  action at  $\beta=5.75$ and  $O(a)$ improved Wilson  (Clover) action
  with  $\kappa=0.1410(0.010)0.1440$, which  roughly covers  the quark
  mass region of $m_s \le  m_q \le 2 m_s$.  $\kappa_s=0.1440$ is fixed
  for $s$ quark, and the  chiral extrapolation is performed by varying
  $\kappa=0.1410-0.1440$ for $u$ and $d$ quarks.
  The  lattice  spacing  is   determined  with  the  Sommer  parameter
  $r_0^{-1}  = 395$  MeV leading  to  the spatial  lattice spacing  as
  $a_s^{-1}  = 1.10$  GeV ($a_s  \simeq 0.18$  fm).
  We  use 504 gauge  configurations for  $J^P=1/2^{\pm}$ penta-quarks,
  1000 gauge configurations for $J^P=3/2^{\pm}$ penta-quarks, and 1872
  gauge  configurations  for   tetra-quarks.  A  smeared  source  with
  gaussian size  $\rho \simeq 0.4$  fm is adopted after  Coulomb gauge
  fixing to suppress the contamination of excited states.

  We use  the hybrid boundary  condition(HBC) \cite{ishii} as  well as
  the  conventional   periodic  BC(PBC)  to   discriminate  a  compact
  resonance  state  from  two-hadron   scattering  states.  HBC  is  a
  flavor-dependent spatial BC,  where we impose PBC on  $s$ quark, and
  anti-PBC  on  $u$  and  $d$ quarks.   Since  $\Theta^+(uudd\bar{s})$
  contains  even number  of $u$  and $d$  quarks, it  obeys  PBC.  Its
  spatial momentum is discretized  as $p_i = 2n_i\pi/L$ ($n_i\in \ZZ$)
  in  the  spatial   box  of  the  size  $L$.    The  5Q  energy  with
  $\vec{p}=\vec{0}$ is  expected to be  insensitive to HBC as  long as
  the  box   can  accommodate  the   5Q  state.  In   contrast,  since
  $N(uud,udd)$ and  $K(u\bar{s}, d\bar{s})$ contain odd  number of $u$
  and  $d$ quarks, they  obey anti-PBC  leading to  the discretization
  $p_i = (2 n_i + 1)\pi/L$.  Since they have the non-vanishing minimum
  momentum $|\vec p_{\rm min}| = \sqrt{3}\pi/L$, the $NK$ threshold is
  raised  as   $E_{NK}=\sqrt{m_N^2  +  3\pi^2/L^2}   +  \sqrt{m_K^2  +
  3\pi^2/L^2}$, which can be used for the discrimination.

  We first consider $J^P=1/2^{\pm}$  iso-scalar 5Q state with a non-NK
  type interpolating field as
  $
    \psi
    \equiv
    \epsilon_{abc}\epsilon_{ade}\epsilon_{bfg}
    ( u_d^T C\gamma_5 d_e )
    ( u_f^T C d_g )
    C\bar{s}_c^T,
    $
  where  $C=\gamma_4\gamma_2$ denotes  the charge  conjugation matrix,
  and $a-g$ the color indices.
  \begin{figure}
    \includegraphics[angle=-90,width=0.45\textwidth]{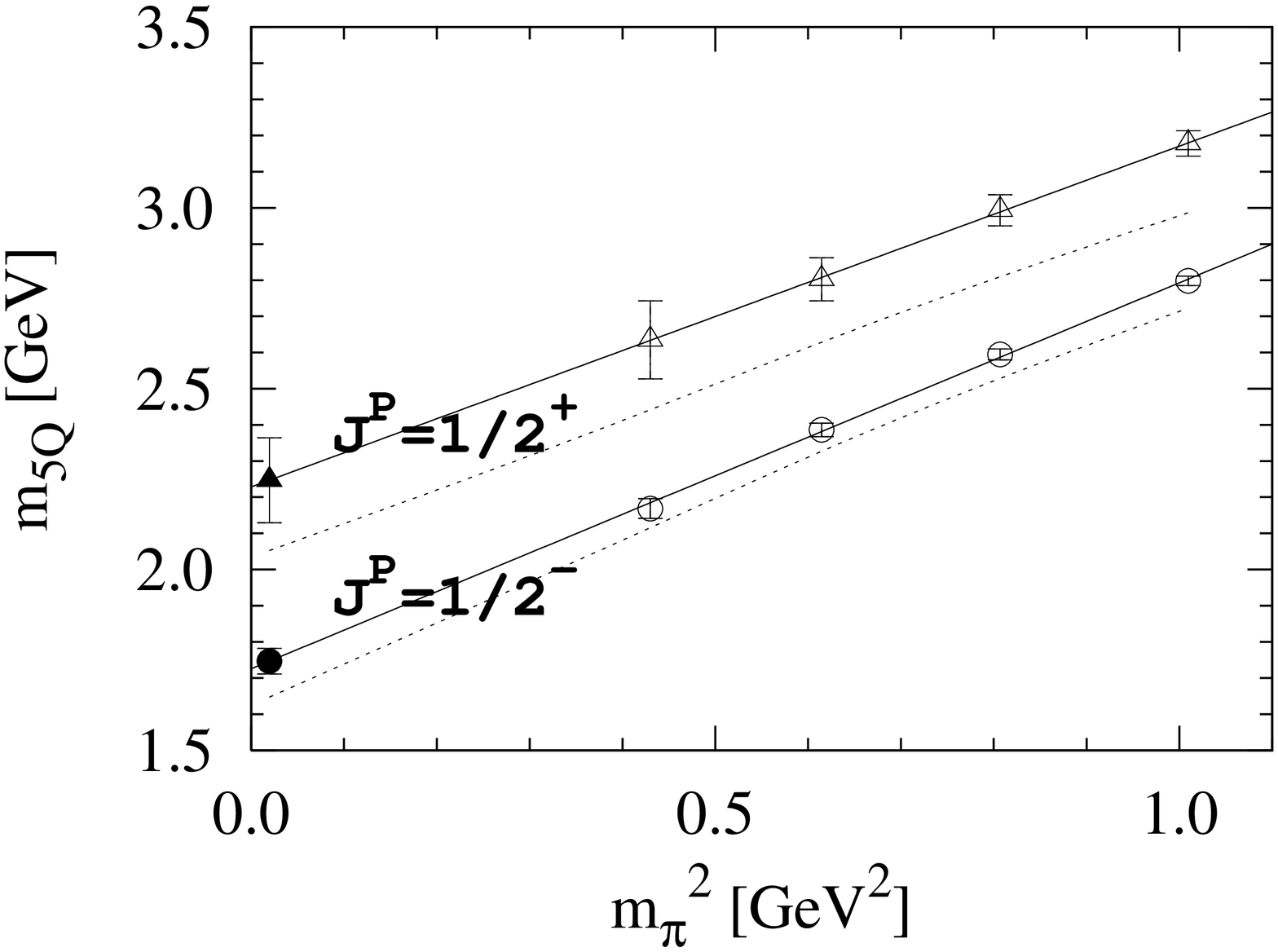}
    \Hs\Hs
    \includegraphics[angle=-90,width=0.45\textwidth]{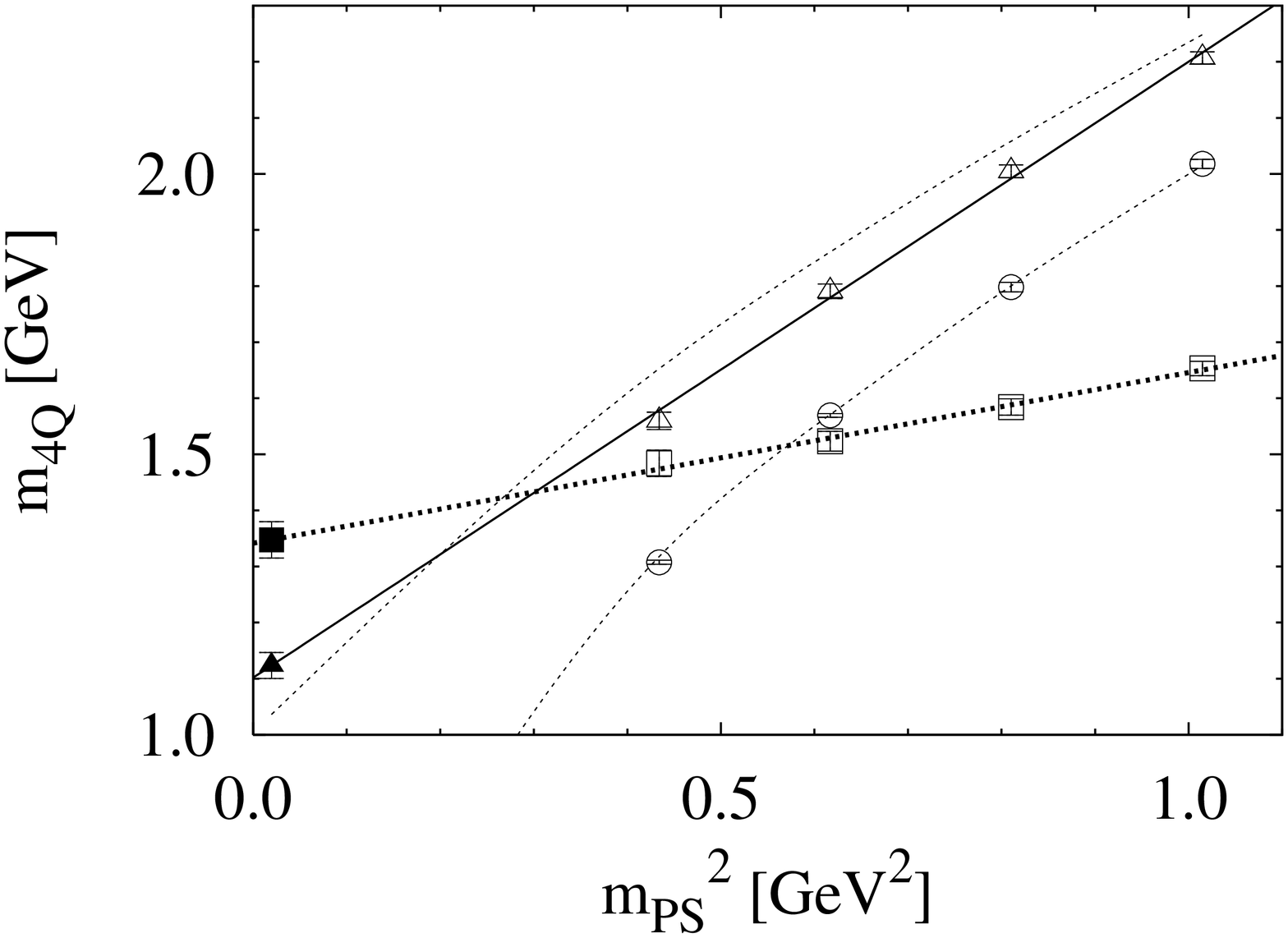}
    \caption{$J^{P}=1/2^{\pm}$   5Q  masses   (left)  and   4Q  masses
    (right). See the main text for explanations.
      \vspace{-2em}
    }
    \label{fig1}
  \end{figure}
  \Fig{fig1}~(left) shows  5Q masses for  $J^P=1/2^{\pm}$.  The dotted
  lines  indicate   the  p-wave(upper)  and   the  s-wave(lower)  $NK$
  thresholds.  The  open symbols denote  the direct lattice  QCD data,
  which are linearly extrapolated  into the physical quark mass region
  denoted by  closed symbols.  We  obtain $m_{5Q} = 2.25(12)$  GeV for
  $J^P=1/2^+$,   which   is   too   massive  to   be   identified   as
  $\Theta^+(1540)$.   For $J^P=1/2^-$,  we obtain  $m_{5Q}  = 1.75(4)$
  GeV,  which  is  located  above  the s-wave  $NK$  threshold(on  the
  lattice)  by about  100 MeV.  Although it  might be  a  candidate of
  $\Theta^+(1540)$ in this sense,
  HBC  analysis  indicates  that   it  is  an  $NK$  scattering  state
  \cite{ishii}.

  We  next consider  $J^P=3/2^{\pm}$ iso-scalar  5Q states  with three
  Rarita-Schwinger interpolating fields as
  $
    \psi_{\mu}
    \equiv
    \epsilon_{abc}\{
    ( u_a^T C\gamma_5 d_b ) u_c
    \cdot
    ( \bar{s}_d \gamma_{\mu} d_d )
    -
    ( u_a^T C\gamma_5 d_b ) d_c
    \cdot
    ( \bar{s}_d \gamma_{\mu} u_d )
    \}
    $,
    $
    \psi_{\mu}
    \equiv
    \epsilon_{abc}\{
    ( u_a^T C\gamma_5 d_b ) u_d
    \cdot
    ( \bar{s}_d \gamma_{\mu} d_c )
    -
    ( u_a^T C\gamma_5 d_b ) d_d
    \cdot
    ( \bar{s}_d \gamma_{\mu} u_c )
    \}
    $,
    and
    $
    \psi_{\mu}
    \equiv
    \epsilon_{abc}\epsilon_{def}\epsilon_{cfg}
    $
    $
    ( u_a^T C\gamma_5 d_b )
    ( u_d^T C\gamma_5 \gamma_{\mu} d_e )
    C \gamma_5 \bar{s}_g
    $.
  Note that $J^P=3/2^-$ assignment  can explain the narrow decay width
  of  $\Theta^+(1540)$  as  pointed  out  by  \Ref{hosaka},  and  that
  $J^P=3/2^+$ assignment  can be supported  by the diquark  picture as
  the LS-partner of $J^P=1/2^+$ 5Q state \cite{jaffe}.  However, after
  the  chiral extrapolations,  we  obtain only  massive  5Q states  as
  $m_{5Q}  \simeq 2.1-2.2$  GeV  for $J^P=3/2^-$,  and $m_{5Q}  \simeq
  2.4-2.6$ GeV for $J^P=3/2^+$. These  5Q states are too massive to be
  identified as $\Theta^+(1540)$ \cite{ishii}.

  $D_{s0}^+(2317)$  is  a  tetra-quark(4Q)  candidate. It  is  so  far
  believed  to  be  iso-scalar  due  to  its  narrow  decay  width  to
  $D_s^+\pi^0$.   Recently,   Terasaki  pointed  out   the  iso-vector
  possibility  \cite{terasaki} in  order to  resolve  the experimental
  constraint in the radiative decay \cite{cleo} as
  $
	{\Gamma(D_{s0}^+(2317) \to D_s^{*+}\gamma)}/
	{\Gamma(D_{s0}^+(2317) \to D_s^{+}\pi^0)}
	<
	0.052.
  $
  If it  is iso-vector,  it has a  manifestly exotic  iso-spin partner
  $D_{s0}^{++}(cu\bar{s}\bar{d})$, which we  will consider here mainly
  in   the   idealized   SU(4)$_f$    limit.    We   employ   a   {\it
  diquark-antidiquark-type} interpolating fields as
  $\phi   \equiv
  \epsilon_{abc}\epsilon_{dec}
  (u_a^T     C\gamma_5     c_b )
  ( \bar{d}_d^T C\gamma_5 \bar{s}_e)$.
  We use a spatial BC similar  to HBC, where we impose anti-PBC on $u$
  and $c$ quarks, and  PBC on $d$ and $s$ quarks. We  will refer to it
  as ``{\it HBC}'' as well.
  So far, we obtain a  positive signal only in the light-quark sector.
  A  preliminary   result  is  shown  in   \Fig{fig1}~(right)  in  the
  idealistic SU(4)$_f$ limit.
  Circles and  triangles denote 4Q  masses obtained with PBC  and HBC,
  respectively.   Thin  dotted  lines denote  two-PS-meson($q\bar{q}$)
  thresholds for PBC(lower) and  HBC(upper).
  Note that 4Q states with PBC(circles) fit the PBC threshold and that
  they  have  no  counterparts  in  HBC  spectrum.   Hence,  they  are
  scattering states.
  In contrast,  4Q states with HBC(triangles) appear  below the raised
  threshold by  about 100 MeV,  and its chiral behavior  is differenct
  from  that   of  the  two-PS-meson   threshold  \cite{ishii},  which
  indicates  the existence of  compact 4Q  resonance state  at $m_{4Q}
  \simeq 1.1$ GeV in the idealized SU(4)$_f$ chiral limit.
  Note  that, if  the contribution  from the  disconnected  diagram is
  negligible,   $D_{s0}^{++}(cu\bar{s}\bar{d})$   is  identical   with
  $f_0(ud\bar{u}\bar{d})$. Then,  this 4Q  state may be  identified as
  one of  the scalar-nonet $f_0(980)$.   For comparison, we  also show
  the scalar  $q\bar{q}$ masses with squares, which  may correspond to
  $f_0(1370)$   or  $a_0(1450)$   consistent   with  the   quark-model
  assignment.
  We  see that  the 4Q  states can  be lighter  than  the conventional
  $q\bar{q}$ in the light quark-mass region.

  To summarize, we have studied $\Theta^+(1540)$ with $J^P=1/2^{\pm}$,
  $3/2^{\pm}$  by  using anisotropic  lattice  QCD.  For  $J^P=1/2^+$,
  $3/2^{\pm}$, we have obtained only massive 5Q states as $m_{5Q} > 2$
  GeV,   which   cannot  be   identified   as  $\Theta^+(1540)$.   For
  $J^P=1/2^-$, we have obtained  a low-lying 5Q state at $m_{5Q}\simeq
  1.75$  GeV, which  has turned  out to  be an  $NK$  scattering state
  through  HBC   analysis  however.   We  have   also  considered  the
  iso-vector  possibility  of   the  4Q  candidate  $D_{s0}^{+}(2317)$
  recently pointed out by \Ref{terasaki}.
  We have obtained a positive signal in the idealized SU(4)$_f$ chiral
  limit at $m_{4Q} \simeq 1.1$ GeV.  Within the valence approximation,
  it    may    correspond    to    one    of    the    scalar    nonet
  $f_0(ud\bar{u}\bar{d})$.  MEM  analysis of 4Q  spectrum is currently
  in progress, which will be presented elsewhere.
  \vspace{-0.1em}

  \begin{center}{\bf ACKNOWLEDGMENTS}\end{center}
  \vspace{-0.4em}
  Lattice QCD calculations have been  performed with NEC SX-5 at Osaka
  University.
\vspace{-1.5em}



\bibliographystyle{aipproc}   

\bibliography{sample}

\IfFileExists{\jobname.bbl}{}
 {\typeout{}
  \typeout{******************************************}
  \typeout{** Please run "bibtex \jobname" to optain}
  \typeout{** the bibliography and then re-run LaTeX}
  \typeout{** twice to fix the references!}
  \typeout{******************************************}
  \typeout{}
 }

\end{document}